\pdfoutput=1

\documentclass[final]{IEEEtran}

\usepackage[utf8]{inputenc} 
\usepackage[T1]{fontenc}    
\usepackage{hyperref}       
\hypersetup{unicode = true, colorlinks,linkcolor=blue,anchorcolor=blue,citecolor=blue,
pdftitle = {Compressed Super-Resolution of Positive Sources},
pdfauthor = {Maxime Ferreira Da Costa, Yuejie Chi}}
\usepackage{url}
\usepackage{algorithmic}
\usepackage{bm,bbm}
\usepackage{amsthm,amssymb,amsmath,mathtools,stmaryrd}
\usepackage{microtype}
\usepackage{xcolor,graphicx,tabularx}
\usepackage{comment}

\renewcommand{\baselinestretch}{0.96}

\theoremstyle{plain}
\newtheorem{thm}{Theorem}

\newtheorem{lem}[thm]{Lemma}

\newtheorem{cor}[thm]{Corollary}
\newtheorem{example}[thm]{Example}

\newcommand{\cA}{{\mathcal{A}}}

\newcommand{\bc}{{\boldsymbol c}}

\newcommand{\bSigma}{{\boldsymbol \Sigma}}
\newcommand{\be}{{\boldsymbol e}}

\newcommand{\bq}{{\boldsymbol q}}

\newcommand{\bu}{{\boldsymbol u}}

\newcommand{\bM}{{\boldsymbol M}}

\newcommand{\bA}{{\boldsymbol A}}

\newcommand{\ba}{{\boldsymbol a}}

\newcommand{\bw}{{\boldsymbol w}}
\newcommand{\bS}{{\boldsymbol S}}
\newcommand{\bx}{{\boldsymbol x}}

\newcommand{\by}{{\boldsymbol y}}

\newcommand{\bH}{{\boldsymbol H}}
\newcommand{\bK}{{\boldsymbol K}}
\newcommand{\bV}{{\boldsymbol V}}

\newcommand{\bP}{{\boldsymbol P}}

\newcommand{\bp}{{\boldsymbol p}}
\newcommand{\bI}{{\boldsymbol I}}

\newcommand{\bbR}{\mathbb{R}}
\newcommand{\bbC}{\mathbb{C}}

\newcommand{\cI}{\mathcal{I}}
\newcommand{\cJ}{\mathcal{J}}
\newcommand{\cT}{\mathcal{T}}

\newcommand\anorm[1]{\left\Vert #1 \right\Vert_{\cA}}
\newcommand\panorm[1]{\left\Vert #1 \right\Vert_{\cA+}}

\renewcommand{\Re}{\operatorname{Re}}

\newcommand{\herm}{\mathsf{H}}

\newcommand{\spann}{\mathop{\rm span}}

\newcommand{\diag}{\mathop{\rm diag}}

\newcommand{\argmin}{\mathop{\rm argmin}}
\newcommand{\argmax}{\mathop{\rm argmax}}

\DeclareMathOperator{\rank}{rank}

\let\tilde\widetilde

\title{Compressed Super-Resolution of Positive Sources}

\author{Maxime Ferreira Da Costa,~\textit{Member, IEEE}, Yuejie Chi,~\textit{Senior Member, IEEE}
\thanks{The authors are with Department of Electrical and Computer Engineering, Carnegie Mellon University, Pittsburgh, PA 15213, USA. Emails: \texttt{\{mferreira,yuejiechi\}@cmu.edu}.}\thanks{This work is supported in part by ONR under the grants N00014-18-1-2142 and N00014-19-1-2404, by NSF under the CAREER grant ECCS-1650449, and by ARO under the grant W911NF-18-1-0303.}}

\begin{document}

\maketitle

\begin{abstract}
Atomic norm minimization is a convex optimization framework to recover point sources from a subset of their low-pass observations, or equivalently the underlying frequencies of a spectrally-sparse signal. When the amplitudes of the sources are positive, a positive atomic norm can be formulated, and exact recovery can be ensured without imposing a separation between the sources, as long as the number of observations is greater than the number of sources. However, the classic formulation of the atomic norm requires to solve a semidefinite program involving a linear matrix inequality of a size on the order of the signal dimension, which can be prohibitive.
In this letter, we introduce a novel ``compressed'' semidefinite program, which involves a linear matrix inequality of a reduced dimension on the order of the number of sources. We guarantee the tightness of this program under certain conditions on the operator involved in the dimensionality reduction. Finally, we apply the proposed method to direction finding over sparse arrays based on second-order statistics and achieve significant computational savings.
\end{abstract}


\section{Introduction}

Super-resolution \cite{candes_super-resolution_2013}\nocite{chi_harnessing_2019}-\cite{candes_towards_2014} is a signal processing problem aiming at recovering point sources from their low-pass observations. It finds broad applications in applied science from the estimation of the direction of arrivals of far-fields signals in classical array processing, to reverting the distortions introduced by the imperfection of the measurement device in modern imaging modalities.

Algorithms based on convex optimization \cite{candes_towards_2014}\nocite{chen_robust_2014}-\cite{tang_compressed_2013} have been recently proposed to solve the super-resolution problem without discretizing the grid \cite{chi_sensitivity_2011}\nocite{herman_high-resolution_2009}-\cite{malioutov_sparse_2005}. Among those, this letter focuses on atomic norm minimization (ANM, a.k.a.~total variation minimization) \cite{tang_compressed_2013}, which proposes to localize the point sources from the output of a semidefinite program (SDP) \cite{mishra_spectral_2015}\nocite{bhaskar_atomic_2013,duval_exact_2015,chi_compressive_2015,li2015off,yang_exact_2016,zhang2019efficient,yang2015enhancing,li2020approximate,duval2020characterization,da2020stable,li2018atomic,fu2018quantized,heckel2016super,heckel_generalized_2018,wang2019super,ongie2016off}-\cite{wei2020gridless}; see \cite{chi_harnessing_2019} for a recent overview. ANM inherits well-established advantages of convex estimators, such as amenability to performance analysis and robustness to the presence of noise. It
is also a versatile  framework that can easily be adapted to fit new measurement models not directly handled by classical approaches \cite{chi_guaranteed_2016}\nocite{li_stable_2019,fernandez-granda_demixing_2017}-\cite{ferreira2019self}. Additionally, ANM is agnostic to the model order. However, the computational complexity of ANM, essentially driven by the size of the SDP, limits its scalability and remains the most prohibitive drawback for practical implantation of this method to real-time systems.

In many imaging applications such as fluorescence microscopy \cite{zhu2012faster}, the point sources are positive, which is a prior that can be leveraged to improve performance \cite{donoho_sparse_2005}\nocite{denoyelle_support_2017,morgenshtern_super-resolution_2016,schiebinger_superresolution_2017,morgenshtern2020super}-\cite{eftekhari2019sparse}. In particular, no separation between positive sources is necessary to guarantee the success of atomic norm minimization as long as the number of observations is greater than the number of sources. In this letter, we propose to solve the super-resolution problem of positive sources using a ``compressed'' ANM algorithm, which only involves a linear matrix inequality (LMI) of dimension on the order of the number of point sources, instead of the signal length. We guarantee  exact reconstruction using the proposed algorithm under certain conditions on the operator involved in the dimensionality reduction, which may lead to significant computational savings. As an illustration, we apply the proposed algorithm in the context of direction finding over sparse arrays from the second-order statistics \cite{pal2010nested}\nocite{qin2015generalized,vaidyanathan2010sparse,qiao2019guaranteed}-\cite{tan2014direction}. Finally, numerical experiments are provided to demonstrate the effectiveness of the proposed algorithm. Our work is related to the compressed ANM proposed in \cite{ferreira_da_costa_low_2017,cosse2020compressed}, but focuses on the positive case where we provide guarantees without imposing any separation condition on the sources.


\section{Problem Formulation and Backgrounds}

Let $\ba(\tau) \in \bbC^N$ be the discrete complex exponential vector
\(
	\ba(\tau) = [1, e^{i 2 \pi \tau}, \dots, e^{i 2\pi (N-1) \tau}]^\top
\)
for $\tau \in [0,1)$. Consider a discrete signal $\bx^\star = [x^\star_0,\dots,x^\star_{N-1}] \in \bbC^N$ modeled as a sparse positive combination of elements of the form $\ba(\tau)$, i.e.
\begin{equation}\label{eq:signal-model}
	\bx^\star = \sum_{k=1}^p c_k^\star \ba(\tau^\star_k),
\end{equation}
for some source locations $\{\tau^\star_k\}_{k=1}^p \subset [0,1)$ and some positive amplitudes $\{c^\star_k\}_{k=1}^p \in \bbR^+$.
The goal of super-resolution is to recover the parameters $\{\tau^\star_k\}$, $\{c^\star_k\}$ from possibly a subset of entries of $\bx^{\star}$ given by $\bx^\star_\Omega = \bP_{\Omega} \bx^\star$, where $\bP_{\Omega} \in \bbC^{\left\vert \Omega \right\vert \times N}$ is the matrix that only retains the entries indexed in the subset $\Omega \subseteq \{0,\dots,N-1\} $. The problem is in general ill-posed, in the sense that there could be infinitely many possible configurations of parameters $\{\tau_k\}$, $\{c_k\}$ that are consistent with the observations. Therefore, it is natural to seek for the {\em sparse decomposition} \eqref{eq:signal-model} that contains the smallest number of point sources.

The atomic norm \cite{chandrasekaran_convex_2012} is a general framework to promote sparse solutions to linear inverse problems. Given a generic atomic set $\cA \subset \mathbb{C}^N$, the atomic norm $\anorm{\bx} \triangleq \inf_{t>0}\{\bx \in t\cA\}$ of a vector $\bx \in \mathbb{C}^N$ is defined by the Minkowski functional of the set $\cA$ evaluated at $\bx$.\footnote{Here, the atomic norm should be interpreted broadly as a pseudo-norm since it may not be a norm when $\cA$ is not centrally symmetric.}  Let $\cT(\bx)$ be the Hermitian Toeplitz matrix whose first column is $\bx$. Specializing the atomic set to the set of \emph{unphased} complex exponential vector, i.e. $\cA+ = \{ \ba(\tau)\, | \, \tau \in [0,1) \} $, the atomic norm simplifies to \cite{chi_harnessing_2019}
\begin{align}\label{eq:positive-atomic-norm}
	\panorm{\bx} ={}& \inf \Big\{ \sum_{k}  c_k   \;\vert \; \bx = \sum_k{ c_k   \ba(\tau_k), \, c_k > 0 } \Big\} \nonumber \\
	={}& \begin{cases}
		\Re( x_0 ) & \text{if } \cT(\bx) \succeq \bm{0} \\
		+ \infty & \text{otherwise}.
	\end{cases}
\end{align}
The decomposition $\bx= \sum_k c_k   \ba(\tau_k)$ that attains the infimum in the first equality of \eqref{eq:positive-atomic-norm} is called the \emph{atomic decomposition}. Of particular interest, it is known that for any vector $\bx_{\star}$ of the form \eqref{eq:signal-model}, as long as $p < N$, its atomic decomposition perfectly recovers the sparse decomposition \eqref{eq:signal-model} and, therefore, provides a means to recover $\{c_k^\star\}$, $\{\tau_k^\star\}$ \cite{de_castro_exact_2012}. The tightness of the atomic decomposition holds without imposing any separation between the positive point sources, which leads to better resolution than the case of signed amplitudes, where a separation is necessary~\cite{ferreira_da_costa_tight_2018,tang2015resolution}.

Given partial observation $\bx^\star_{\Omega}$, one can recover the ground truth signal $\bx^\star$ by solving the ANM problem as
\begin{align}\label{eq:SCE}
&	\widehat{\bx}_{\textsf{ANM}} :={} \argmin_{\bx \in \bbC^N} \; \panorm{\bx} \mbox{ s.t. } \bx_{\Omega} = \bx^\star_{\Omega} \nonumber \\
&	\;={} \argmin_{\bx \in \bbC^N} \; \Re( x_0 ) \mbox{ s.t. } \bx_{\Omega} = \bx^\star_{\Omega} \mbox{ and } \cT(\bx) \succeq \bm{0}. \tag{ANM}
\end{align}
This approach is guaranteed to yield a perfect reconstruction of $\bx^\star$ as long as $p < \left\vert \Omega \right\vert$, without any need for randomness of the observation set $\Omega$ \cite{eftekhari2019sparse}.
However, from the computational perspective, \eqref{eq:SCE} involves an LMI of dimension $N$, which in practice may be prohibitive to solve.

\section{Main Results}

Inspired by \cite{ferreira_da_costa_low_2017,cosse2020compressed}, we propose a novel approach to reduce the computational complexity of \eqref{eq:SCE} by projecting the positivity constraint $\cT(\bx) \succeq \bm{0}$ to a lower dimension. Consider a matrix $\bM \in \bbC^{M \times N}$ with $M \leq N$ and full rank, i.e. $\rank(\bM) = M$. The compressed positive ANM program is given by
\begin{align}\label{eq:C-SCE}
	\widehat{\bx}_{\textsf{C-ANM}} :={}& \argmin_{\bx \in \bbC^N} \; \Re( x_0 ) \nonumber\\
	 \mbox{  s.t.  } \quad& \bx_{\Omega} = \bx^\star_{\Omega}  \mbox{ and } \bM \cT(\bx) \bM^\herm \succeq \bm{0}. \tag{C-ANM}
\end{align}
Note that the compressed SDP \eqref{eq:C-SCE} now contains an LMI of dimension $M \leq N$. An immediate question arising from the definition of \eqref{eq:C-SCE} concerns its tightness, i.e. the conditions under which its solution uniquely recovers $\bx^\star$ and its sparse decomposition. Similar to many linear inverse problems, the tightness of \eqref{eq:C-SCE} is related to the existence of a so-called \emph{dual certificate}: an element lying in the dual feasible set and attaining the optimum of the cost function. Lemma \ref{lem:dualCertificate} characterizes the dual certificate conditions that certify the tightness of \eqref{eq:C-SCE}, whose proof is given in Appendix \ref{sec:ProofOfDualCertificate}.

\begin{lem}[Dual certificate]\label{lem:dualCertificate}
	Suppose there exists a trigonometric polynomial $Q(\tau) := \sum_{n=0}^{N-1} q_n e^{-i 2 \pi n \tau}$ with coefficient vector $\bq = [q_0, \dots, q_{N-1}]^\top \in \bbC^N$ such that
	\begin{enumerate}
		\item $Q(\tau)\neq 1$.
		\item The equality $1 - \Re (Q(\tau)) = 0$ holds for $\tau \in \{\tau^\star_k  \}_{k=1}^p$.
		\item The coefficient vector $\bq$ verifies $\bq_{\Omega^c} = \bm{0}$.
		\item
		There exists a countable collection of trigonometric polynomials $\{P_{i}(t) \}_i$ with coefficients $\bp_i \in \spann(\bM^\herm)$ such that
		\(
			1 - \Re (Q(\tau)) = \sum_{i} \left\vert P_i(\tau) \right\vert^2 \geq 0
		\) for all $\tau \in [0,1)$.
	\end{enumerate}
	Then, $\widehat{\bx}_{\textsf{C-ANM}} = \bx^\star$ is the unique solution of \eqref{eq:C-SCE}.
\end{lem}

From Lemma \ref{lem:dualCertificate}, it suffices to construct a trigonometric polynomial $Q(\tau)$ verifying specific conditions to conclude on the tightness of the compressed SDP \eqref{eq:C-SCE}. The essential difference between the conditions for \eqref{eq:C-SCE} stated in Lemma \ref{lem:dualCertificate}, and those for \eqref{eq:SCE} \cite{tang_compressed_2013} on the dual certificate is in the fourth assumption. Herein, the polynomial $1-\Re(Q(\tau))$ must have a sum-of-squares (SOS) structure over a low-dimensional subspace of trigonometric polynomials whose coefficients lie in the span of $\bM^\herm$, also called the {\em sparse-SOS} condition. Note that, in the absence of compression (i.e. $\bM = \bI$), the sparse-SOS assumption always holds as a consequence of the Fej\'er-Riesz theorem \cite{fejer1916trigonometrische}.

A natural question is then: how to design the matrix $\bM$ that verifies Lemma \ref{lem:dualCertificate}? In the sequel, we focus on a specific design and discuss its tightness.
Let $\cI$ be a subset of $\{0,\dots, N-1\}$ with $\{0\}\in \cI$ and cardinality $M = \left\vert \cI \right\vert$, and we denote
by $\partial\cI$ the set of the positive pairwise differences of elements in $\cI$, given by
\begin{equation}\label{eq:difference-set}
	\partial\cI = \left\{ j\;|\; j = i_1 - i_2 \geq 0, \;(i_1,i_2)\in \cI \times \cI \right\}.
\end{equation}
Let $\bM = \bP_\cI \in \bbC^{M \times N}$ be the subsampling matrix selecting the elements whose indices belong to $\cI$. Theorem \ref{thm:exact-reconstruction} states that with this choice of $\bM$, \eqref{eq:C-SCE} is tight if $\partial \cI \subseteq \Omega$ and the number of sources $p<M$. The proof is given in Appendix~\ref{sec:proof-exact-reconstruction}.
\begin{thm}[Exact reconstruction]\label{thm:exact-reconstruction}
	Let  $\cI \subseteq \{0,\dots, N-1\}$ be a subset of cardinality $M = \left\vert \cI \right\vert$ verifying $0\in \cI$ and $\partial \cI \subseteq \Omega$. Moreover, suppose that $p < M $. Then for the choice $\bM = \bP_\cI$, $\widehat{\bx}_{\textsf{C-ANM}} = \bx^\star$ is the unique solution of~\eqref{eq:C-SCE}.
\end{thm}

Theorem \ref{thm:exact-reconstruction} suggests that the computational complexity can be significantly reduced, where the LMI has a dimension on the order of the number of sources $p$. For example, consider the case will full observation, i.e. $\Omega=\{0,\ldots,N-1\}$. Then,
Theorem \ref{thm:exact-reconstruction} guarantees the compressed ANM \eqref{eq:C-SCE} is exact for any $\cI$ as long as $|\cI|>p$ and $0\in \cI$. As another example, when $\Omega=\{0,\ldots,p\}$, choosing $\cI = \Omega$ also reduces the complexity significantly to the order of $p$.

\section{Application: Direction Finding in Sparse Arrays}

In this section, we illustrate the applicability of Theorem~\ref{thm:exact-reconstruction} for direction finding in sparse arrays from second-order statistics \cite{pal2010nested,tan2014direction}, which is a problem of great interest in the array processing literature, as this approach allows the recovery of more sources than the number of antennas, and offers better resolution than its first-order counterpart. We show in particular that \eqref{eq:C-SCE} can be applied to reduce the computational complexity of ANM-based recovery, where \eqref{eq:C-SCE} can recover the sources by involving an LMI of the size equal to the number of sources, instead of the size of the aperture.

\subsection{Exact Recovery with Infinite Snapshots }
We start by introducing some notation. Let $\cJ$ be the set of integer indices corresponding to the location of the antennas in a linear sparse array. The aperture of $\cJ$ is assumed to be $N$, so that $\cJ$ can be embedded in a uniform array of $N$ elements, i.e., $\cJ \subseteq \{0,\dots,N-1\}$ and $\{0,N-1\} \in \cJ$. We denote by $\Omega = \partial \cJ$ the set of the positive indices of the difference co-array, where $\partial \cJ$ is given as in \eqref{eq:difference-set}.

At the time instance $\ell = 1,\dots,L$, the noiseless received signal $\bu^\star_\ell \in \bbC^{M}$ is modeled as
\(
	\bu_\ell^\star = \bP_\cJ \sum_{k=1}^p c_{k,\ell}^\star \ba(\tau_k^\star) + \bw_\ell,
\)
for some zero-mean $c_{k,\ell}^{\star} \in \mathbb{C}$, and $\bw_\ell \in \mathbb{C}^{|\cJ|}$ is a white additive noise with zero-mean and variance $\sigma^2$, which is assumed known. If the sources are  incoherent, i.e.
 obey the second-order statistical property:
 \begin{align}\label{second_order}
 \mathbb{E}\left[\overline{c_{k,\ell}^{\star}} c_{k^{\prime},\ell^{\prime}}^{\star}\right] = \begin{cases}
 \eta_k^2, & \mbox{if}~ k = k^{\prime},\; \ell = \ell^{\prime} \\
 0, & \mbox{otherwise}
 \end{cases},
 \end{align}
then the covariance matrix $\bm{\Sigma}^\star_\cJ = \mathbb{E}[\bu_\ell^\star (\bu_\ell^\star)^{\mathsf{H}} ]  $ writes
 \begin{equation} \label{eq:cov_obs}
 \bm{\Sigma}^{\star}_\cJ
 = \bP_\cJ \bSigma^\star \bP_\cJ^\herm + \sigma^2 \bI_{\left\vert \cJ \right\vert},
 \end{equation}
 where $\bSigma^\star = \sum_{k=1}^p \eta_k^2 \ba(\tau_k) \ba(\tau_k)^\herm$ is a positive semidefinite Hermitian Toeplitz matrix corresponding the covariance of the observations gathered on the full uniform array $\{0,\dots,N-1\}$, and $\bI_{\left\vert \cJ \right\vert}$ is an identity matrix of size $|\cJ|$. Denote by $\bx^\star \in \mathbb{C}^N$ the first column of $\bSigma^\star$, where
\(
	\bm{x}^{\star} = \sum_{k=1}^p \eta_k^2 \ba(\tau^\star_k)
\)
is a sparse positive linear combination of $p$ discrete complex exponentials, with frequencies $\{\tau^\star_k \}$ encoding the location of the sources. From \eqref{eq:cov_obs}, identifying and rearranging the entries of $\bm{\Sigma}^{\star}_\cJ- \sigma^2 \bI_{\left\vert \cJ \right\vert}$ in the coordinates $(i_1,i_2)\in \partial \cJ$ then equivalently give the observation model $\bx_{\Omega}^{\star}= \bP_{\Omega} \bx^\star$. Theorem \ref{thm:exact-reconstruction} then applies, and the sources can be exactly recovered by applying \eqref{eq:C-SCE} on the vector $\bx_{\Omega}^\star $. This yields the following corollary.

\begin{cor}[Exact recovery over sparse arrays] \label{cor:sparseArray}
	If $\cJ  \subseteq \{0,\dots,N-1 \}$ is a sparse array with $M = \left\vert \cJ \right\vert$ elements and  $\{0,N-1\} \in \cJ$. If $p < M $, then  $\{ \tau^\star_1, \dots, \tau^\star_p \}$ can be exactly recovered using \eqref{eq:C-SCE} with $\bM = \bm{P}_{\cJ}$.
\end{cor}
Corollary \ref{cor:sparseArray} ensures that \eqref{eq:C-SCE} returns the sources by solving an SDP with  LMI of size $M = \left\vert \cJ \right\vert$, while proceeding to the full-dimensional ANM \eqref{eq:SCE} would require to solve an SDP with  LMI of size $N$, equal to the length of the full array. Therefore, the proposed compressed approach can bring an order-of-magnitude reduction in the computational complexity of the problem for appropriate choices of the array and the compression operator. As an example, the benefits are well highlighted when considering the Cantor arrays, which are complete sparse arrays constructed through a fractal process, see \cite{liu2017maximally} for an introduction.
\begin{example}[Cantor arrays]
	If $\cJ \subseteq \{0,\dots,N-1\}$ is a Cantor set, and if $p < \left\vert \cJ \right\vert = M $, then we can recover the spikes by solving a semidefinite program involving a linear matrix inequality of dimension $M = N^{\log(2)/\log(3)} \simeq  N^{0.62}$. Experimental runtimes of algorithms \eqref{eq:SCE} and \eqref{eq:C-SCE} are compared in Table \ref{tab:runtime} for different size of Cantor sets.
\end{example}

In addition, it is worth to pay attention to a particular category of sparse arrays, which are called \emph{complete}, that the difference co-array has no holes, i.e. $\partial \cJ = \{0,\dots,N-1\}$. In that case, Theorem \ref{thm:exact-reconstruction} guarantee that running \eqref{eq:C-SCE} with any compression matrix $\bM = \bP_\cI$ such that $\cJ \subseteq \cI$ would guarantee an exact recovery of at most $\left\vert \cI \right\vert - 1$ sources. Hence, there is a trade-off between the compression ratio of the LMI in \eqref{eq:C-SCE} and the number of sources that can effectively be recovered.

\subsection{Recovery under a Finite Number of Snapshots}

In practice, the exact covariance $\bm{\Sigma}_\cJ^\star$ in \eqref{eq:cov_obs} is imperfectly known, as the number of snapshots $L$ is finite. The empirical covariance of the received signals $\bSigma_\cJ = \frac{1}{L} \sum_{\ell=1}^L \bu_\ell^\star (\bu_\ell^\star)^{\mathsf{H}}$ provides a more accurate estimate of $\bm{\Sigma}_\cJ^\star$ as $L$ increases. Denote by $\by_\Omega \in \mathbb{C}^{\left\vert \cJ \right\vert}$ the noisy estimate of $\bx^\star_\Omega$ obtained from $\bm{\Sigma}_\cJ - \sigma^2 \bI_{|\cJ|}$ in a similar manner as earlier. To adapt to this uncertainty, we formulate the atomic norm denoiser \cite{bhaskar_atomic_2013} by adding a data fidelity term to the cost function of \eqref{eq:C-SCE},
\begin{align}\label{eq:denoiser}
	\widehat{\bx}_{\lambda} :={} \argmin_{\bx \in \bbC^N} \;{}& \frac{1}{2} \left\Vert \bx_\Omega   -  \by_\Omega \right\Vert_2^2 + \lambda  \Re( x_0 ) \nonumber\\
	\mbox{s.t.}\quad{}& \bM \cT(\bx) \bM^\herm \succeq \bm{0}. \tag{C-ANM-Noisy}
\end{align}
where $\lambda > 0$ is a regularization parameter.

\begin{table}[t]
	\centering
	\begin{tabular}{c|c|c|c|c}
	Cantor \# & Aperture& \# elements&  \eqref{eq:SCE} (s)&  \eqref{eq:C-SCE} (s)\\ \hline
	\hline
	$3$ & $10$  & $2^3 = 8$ & $0.30$ & $0.26$  \\  \hline
	$4$ & $28$ & $2^4 = 16$ & $0.34$ & $0.26$  \\  \hline
	$5$ & $82$ & $2^5 = 32$ & $0.66$ & $0.29$  \\  \hline
	$6$ & $244$ & $2^6 = 64$ & $5.83$ & $0.87$  \\  \hline
	$7$ & $730$ & $2^7 = 128$ & $135.62$ & $8.38$  \\  \hline

	\end{tabular}
	\vspace{0.03in}
	\caption{\label{tab:runtime} Runtimes of the \eqref{eq:SCE} and \eqref{eq:C-SCE} algorithms over Cantor arrays of different orders. Both algorithms are implemented via Matlab using the Mosek solver in CVX \cite{grant_cvx:_nodate}. We fix $p = 8$ in the experiments, the results are averaged over 50 trials. \vspace{-0.2in}}
	\end{table}

	\begin{figure}[t]
		\centering
		\includegraphics[width=0.95\columnwidth]{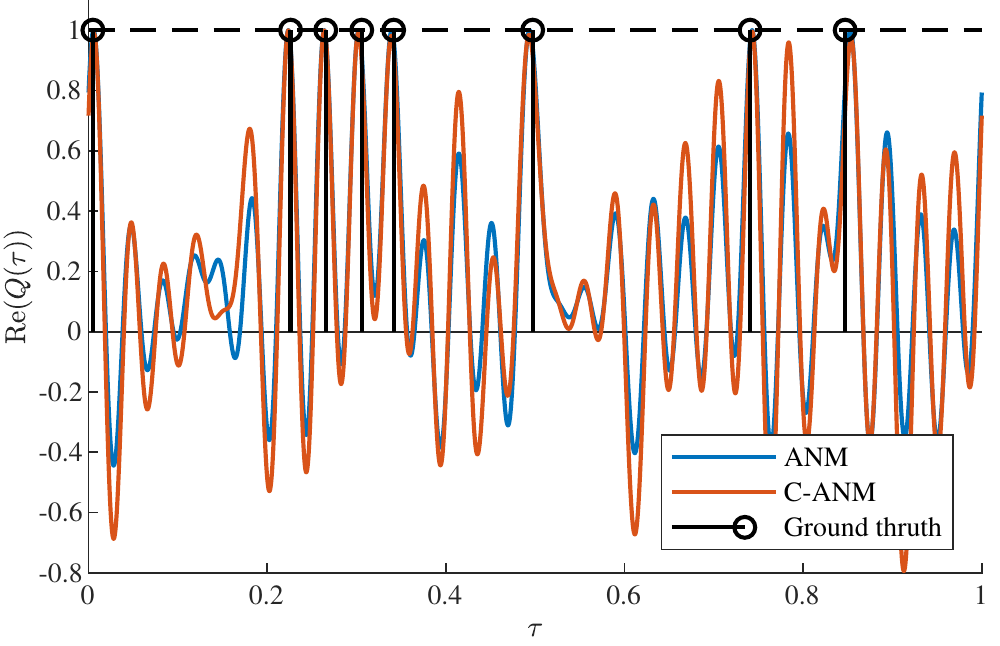}
		\caption{Empirical localization of the sources using a Cantor array from the second order statistics. The picks of the dual polynomials returned by \eqref{eq:denoiser} when $\bM = \bI$ (ANM), and $\bM = \bP_\cJ$ (C-ANM) provide an estimate of the ground truth sources. In those settings, $N=28$, $\left\vert \cJ \right\vert = 16$, $p=8$. We take $L=100$ snapshots, and SNR is $-5$dB.}
		\label{fig:tradeoff}
	\end{figure}

In Figure~\ref{fig:tradeoff}, we compare the localization of the sources using a sparse array from  $L=100$ snapshots using \eqref{eq:denoiser} in the absence of compression ($\bM = \bI$) which corresponds to the original ANM method, and using a compression matrix $\bM = \bP_\cJ$. We pick a Cantor array with aperture $N=28$ and $M=16$ elements. The ground truth signal $\bx^\star$ impinging on the array is formed by $p=8$ incoherent sources with equal unit power $\eta_k^2 = 1$. The signal-to-noise ratio (SNR) is defined as $\mathsf{SNR} = \sum_{k=1}^p \eta_k^2 / \sigma^2$ and is set to $-5$dB. The sources are identified from the dual solution of \eqref{eq:denoiser}, namely, by identifying the peaks of the dual polynomial $\Re \left(Q(\tau) \right) = \Re \left(  \ba(\tau)^{\mathsf{H}} \bq   \right)$, where $\bq \in \mathbb{C}^N$ is the solution to the dual program of \eqref{eq:denoiser} \cite{chi_harnessing_2019}. It can be seen that the compressed ANM approach is able to recover the direction-of-arrivals at a much lower computational complexity.

\section{Conclusions}
In this letter, we showed that super-resolution of point sources can be solved using a compressed ANM algorithm with much lower computational complexity, provided an application of this results to direction finding over sparse arrays. For future work, we aim to study the performance of the compressed ANM in the noisy setting, where it is expected that the compression leads to interesting statistical-computational trade-offs \cite{chandrasekaran2013computational}.

\appendix

\subsection{Proof of Lemma \ref{lem:dualCertificate}} \label{sec:ProofOfDualCertificate}

The Lagrange dual program of \eqref{eq:C-SCE} reads
\begin{align}\label{eq:C-SCE-dual}
&\argmax_{\bq \in \bbC^{N}, \bS \in \bbC^{M \times M}}  \; \Re \left\langle \bx_\Omega, \bq_\Omega \right\rangle  \\
	\mbox{s.t. } &\quad \bS \succeq 0, \quad \bq_{\Omega^c} = \bm{0}, \quad \cT^\ast(\bM^\herm \bS \bM) + \bm{K} \bq = \bm{e}_0,  \nonumber
\end{align}
where $\cT^\ast : \bbC^{N\times N} \to \bbC^{N}$ is the adjoint to the Toeplitz Hermitian operator $\cT$ and is given for any $\bH$ by
\(
	\cT^\ast(\bH)_j = \sum_{\ell=1}^{N-j} H_{j+\ell,\ell},
\)
and $\bm{K} \in \bbC^{N\times N}$ is the diagonal matrix
\(
	\bm{K} = \diag(1,\frac{1}{2},\dots, \frac{1}{2} ).
\)
Suppose there exists $\bq$ that verifies the hypotheses of Lemma \ref{lem:dualCertificate}. Since $\bp_i \in \spann(\bM^\herm)$, there exists $ \bu_i \in  \bbC^M $ such that $\bp_i = \bM^\herm \bu_i$ for all $i$.

\paragraph{Tightness} We start by showing that $\bx^\star$ is a solution to \eqref{eq:C-SCE}. First, as $Q(\tau)\neq 1$, there exists some $i$ such that $P_i(\tau)\neq 0$, and consequently $\bp_i \neq \bm{0}$. Since $\bM$ is full rank by assumption, this implies that $\bu_i \neq \bm{0}$ for some $i$.

Next, the equality $1 - \Re(Q(\tau)) = \sum_{i} \left\vert P_i(\tau) \right\vert^2$ holds for any $\tau \in [0,1)$ if and only if
\begin{align}
	 \be_0 &- \bK \bq ={} \cT^\ast \Big(\sum_i \bp_i \bp_i^\herm \Big) = \cT^\ast \Big(\sum_i \bM^\herm \bu_i \bu_i^\herm \bM \Big)\nonumber \\
	& ={} \cT^\ast \Big(\bM^\herm \Big( \sum_i  \bu_i \bu_i^\herm \Big)  \bM \Big) = \cT^\ast \Big(\bM^\herm \bS  \bM \Big),
\end{align}
where we let $\bS = \sum_{i} \bu_i \bu_i^\herm \succeq 0$ in the last equality. As $\bq_{\Omega^c} = \bm{0}$ by assumption, it is evident that the pair $(\bq, \bS)$ is in the feasible set of the dual problem \eqref{eq:C-SCE-dual}. Evaluating the dual cost function at  $(\bq, \bS)$ yields
\begin{align}
	\Re \left\langle \bx_\Omega, \bq_\Omega \right\rangle =& \Re \left\langle \bx^\star_\Omega, \bq_\Omega \right\rangle
	={} \Re \left\langle \bx^\star, \bq \right\rangle \nonumber \\
	={}& \sum_{k=1}^p c^\star_k \Re \left( \ba(\tau^\star_k)^\herm \bq \right)
	={} \sum_{k=1}^p c^\star_k \Re \left( Q(\tau^\star_k) \right) \nonumber \\
	={}& \sum_{k=1}^p c^\star_k
	={} \panorm{\bx^\star}.
\end{align}
By strong duality, $\bx^\star$ is a solution of \eqref{eq:C-SCE}.

\paragraph{Uniqueness} We now prove that $\bx^\star$ is the unique solution to \eqref{eq:C-SCE}. Suppose that $\widetilde{\bx} \in \bbC^N$ is a solution to \eqref{eq:C-SCE}, and let $\widetilde{\bx} = \sum_{k=1}^{\widetilde{p}} \widetilde{c}_k \ba(\widetilde{\tau}_k)$ an atomic decomposition with $\panorm{\widetilde{\bx}} = \sum_{k=1}^{\widetilde{p}} \widetilde{c}_k$. Since $\tilde{\bx}$ is a solution, we also have that $\panorm{\widetilde{\bx}} = \panorm{\bx^\star}$ and $\widetilde{\bx}_{\Omega} = \bx^\star_\Omega$.

Denote by $R \subset [0,1)$ the set of the roots to the equation $1 - \Re (Q(\tau)) = 0$. As $Q(\tau)$ is not the constant polynomial equal to one, the set $R$ is finite and we have $\left\vert R \right\vert \leq N-1$. By strong duality, we can further write
\begin{align}
	\sum_{k=1}^{\widetilde{p}} \widetilde{c}_k ={}& \panorm{\widetilde{\bx}} ={} \Re \left\langle \bx_\Omega, \bq_\Omega \right\rangle ={} \Re \left\langle \widetilde{\bx}_\Omega, \bq_\Omega \right\rangle
	={} \Re \left\langle \widetilde{\bx}, \bq \right\rangle \nonumber \\
	={}& \sum_{k=1}^{\widetilde{p}} \widetilde{c}_k \Re \left( \ba(\widetilde{\tau}_k)^\herm \bq \right) ={} \sum_{k=1}^{\widetilde{p}} \widetilde{c}_k \Re \left( Q(\widetilde{\tau}_k) \right).
\end{align}
We conclude using the positivity of the $\widetilde{c}_k$s that $\Re (Q(\widetilde{\tau}_k)) = 1$ for $k=1,\dots \widetilde{p}$, and therefore that $\{ \widetilde{\tau}_k \}_{k=1}^{\widetilde{p}} \subseteq R$. Let $\bV_\Omega \in \bbC^{\left\vert \Omega \right\vert \times \left\vert R \right\vert}$ be the matrix whose column are elements of the form $\ba(\tau)$ with $\tau \in R$. We have that $ \bx^\star_\Omega =\bV_\Omega \bc^\star$ and $\widetilde{\bx} = \bV_\Omega \widetilde{\bc}$ for some $\bc^\star$, $\widetilde{\bc}\in \bbC^{\left\vert R \right\vert}$ with $\left\Vert \bc^\star \right\Vert_1 = \left\Vert \widetilde{\bc} \right\Vert_1 = \panorm{\bx^\star}$. We conclude using the uniqueness of the solution to the positive linear program \cite{donoho_sparse_2005}
\begin{align}
	\widehat{\bc} := \argmin_{\bc \in \bbC^{\left\vert R \right\vert}} & \; \left\Vert \bc \right\Vert_1 \mbox{ such that }  \bc \geq 0 \mbox{ and } \bV_\Omega \bc = \bx^\star_\Omega,
\end{align}
that $\widehat{\bc} = \bc^\star = \widetilde{\bc}$, and consequently that $\widetilde{\bx} = \bx^\star$. We conclude that $\bx^\star$ is the unique solution to \eqref{eq:C-SCE}. \qed

\subsection{Proof of Theorem \ref{thm:exact-reconstruction}}\label{sec:proof-exact-reconstruction}

In view of Lemma~\ref{lem:dualCertificate}, it suffices to show the existence of a trigonometric polynomial $Q(\tau)$ verifying the conditions of Lemma~\ref{lem:dualCertificate} for the compression matrix $\bM = \bP_\cI$.

Denote the Vandermonde matrix $\bA= [\ba(\tau_1),\dots,\ba(\tau_p)]\in \bbC^{N \times p}$. As long as $p < M$, the matrix $\bV = \bA^\herm \bP_\cI^\herm \in \bbC^{p \times M}$ has a non-trivial nullspace. Denote by $\bu \in \ker(\bV)$ a non-zero element of this nullspace. We have that
\[
	\bV \bu = \bA^\herm \bP_\cI^\herm \bu = \bm{0}.
\]
Let $\bp = \bP_\cI^\herm \bu$, and $P(\tau)$ be the trigonometric polynomial $P(\tau) = \sum_{n=0}^{N-1} p_k e^{-i 2 \pi n t}$. Moreover, let $Q(\tau) = \sum_{n=0}^{N-1} q_k e^{i 2 \pi n t}$ be such that
\begin{equation}\label{eq:qRelation}
	1 - \Re(Q(\tau)) = \left\vert P(\tau) \right\vert^2, \quad \forall \tau \in [0,1),
\end{equation}
which holds if and only if the vector $\bq \in \bbC^N$ satisfies
\begin{equation}\label{eq:qCoefficients}
	\be_0 - \bK \bq = \cT^\ast(\bp \bp^\herm).
\end{equation}

We now verify that $Q(\tau)$ meets the conditions of Lemma \ref{lem:dualCertificate}.
\begin{enumerate}
\item Since $\bV \be_0 = \bA^\herm \bP_\cI^\herm \be_0 = \bA^\herm \be_0 \neq \bm{0}$, which follows by $0 \in \partial I$, the vector $\bu$ is not collinear to $\be_0$. Thus $\bp$ is not collinear to $\be_0$ and as $\bK$ is a diagonal matrix, from \eqref{eq:qCoefficients} we have
\begin{equation}\label{eq:bp-expression}
	\bq = \bK^{-1} \left(\be_0 - \cT^\ast(\bp \bp^\herm)\right),
\end{equation}
is also not collinear to $\be_0$. It follows that $Q(\tau)\neq 1$.
\item By the assumption $\bu \in \ker(\bV)$, we have that for all $k=1,\ldots, p$,
\begin{align*}
	1 - \Re(Q(\tau^\star_k)) =& \left\vert P(\tau^\star_k) \right\vert^2  \\
													={}& \ba(\tau^\star_k)^\herm \bp \bp^\herm \ba(\tau^\star_k) \\
={}& \ba(\tau^\star_k)^\herm \bP_\cI^\herm \bu \bu^\herm \bP_\cI \ba(\tau^\star_k) \\
={}& \be_k^\herm \bV \bu \bu^\herm \bV^\herm \be_k = 0.
\end{align*}

\item As $\bp \in \spann(\bP_\cI^\herm)$ is supported over $\cI$, the vector $\cT^\ast(\bp \bp^\herm)$ is supported over $\partial \cI$. Since $0 \in \partial I$, the vector $\be_0$, and the difference $\cT^\ast(\bp \bp^\herm) - \be_0$ are also supported over $\partial \cI$.  Since $\bm{K}^{-1}$ is a diagonal matrix, it leaves the support of the subvectors invariant by multiplication. By \eqref{eq:bp-expression}, $\bq$ is supported over $\partial \cI$. By the assumption of Theorem \ref{thm:exact-reconstruction}, we have $\partial \cI \subseteq \Omega$, hence $\Omega^c \subseteq {\partial\cI}^c$, and we conclude that $\bq_{\Omega^c} = 0$.
\item Finally, \eqref{eq:qRelation} holds by construction with $\bp \in \spann(\bP_\cI^\herm)$, thus $Q(\tau)$ is sparse-SOS over $\spann(\bP_\cI^\herm)$.
\end{enumerate}
Invoking Lemma \ref{lem:dualCertificate} concludes the proof of Theorem \ref{thm:exact-reconstruction}. \qed

\bibliography{Atomic.bib}

\begin{thebibliography}{10}
\providecommand{\url}[1]{#1}
\csname url@samestyle\endcsname
\providecommand{\newblock}{\relax}
\providecommand{\bibinfo}[2]{#2}
\providecommand{\BIBentrySTDinterwordspacing}{\spaceskip=0pt\relax}
\providecommand{\BIBentryALTinterwordstretchfactor}{4}
\providecommand{\BIBentryALTinterwordspacing}{\spaceskip=\fontdimen2\font plus
\BIBentryALTinterwordstretchfactor\fontdimen3\font minus
  \fontdimen4\font\relax}
\providecommand{\BIBforeignlanguage}[2]{{%
\expandafter\ifx\csname l@#1\endcsname\relax
\typeout{** WARNING: IEEEtran.bst: No hyphenation pattern has been}%
\typeout{** loaded for the language `#1'. Using the pattern for}%
\typeout{** the default language instead.}%
\else
\language=\csname l@#1\endcsname
\fi
#2}}
\providecommand{\BIBdecl}{\relax}
\BIBdecl

\bibitem{candes_super-resolution_2013}
E.~J. Cand{\`e}s and C.~Fernandez-Granda, ``Super-resolution from noisy data,''
  \emph{Journal of Fourier Analysis and Applications}, vol.~19, no.~6, pp.
  1229--1254, 2013.

\bibitem{chi_harnessing_2019}
Y.~{Chi} and M.~{Ferreira Da Costa}, ``Harnessing sparsity over the continuum:
  Atomic norm minimization for superresolution,'' \emph{IEEE Signal Processing
  Magazine}, vol.~37, no.~2, pp. 39--57, March 2020.

\bibitem{candes_towards_2014}
E.~J. Cand{\`e}s and C.~Fernandez-Granda, ``Towards a mathematical theory of
  super-resolution,'' \emph{Communications on Pure and Applied Mathematics},
  vol.~67, no.~6, pp. 906--956, 2014.

\bibitem{chen_robust_2014}
Y.~Chen and Y.~Chi, ``Robust {Spectral} {Compressed} {Sensing} via {Structured}
  {Matrix} {Completion},'' \emph{IEEE Transactions on Information Theory},
  vol.~60, no.~10, pp. 6576--6601, 2014.

\bibitem{tang_compressed_2013}
G.~Tang, B.~N. Bhaskar, P.~Shah, and B.~Recht, ``Compressed sensing off the
  grid,'' \emph{IEEE Transactions On Information Theory}, vol.~59, no.~11, pp.
  7465--7490, 2013.

\bibitem{chi_sensitivity_2011}
Y.~Chi, L.~L. Scharf, A.~Pezeshki, and A.~R. Calderbank, ``Sensitivity to basis
  mismatch in compressed sensing,'' \emph{IEEE Transactions on Signal
  Processing}, vol.~59, no.~5, pp. 2182--2195, 2011.

\bibitem{herman_high-resolution_2009}
M.~Herman and T.~Strohmer, ``High-resolution radar via compressed sensing,''
  \emph{IEEE Transactions on Signal Processing}, vol.~57, no.~6, pp.
  2275--2284, Jun. 2009.

\bibitem{malioutov_sparse_2005}
D.~Malioutov, M.~{\c C}etin, and A.~S. Willsky, ``A sparse signal
  reconstruction perspective for source localization with sensor arrays,''
  \emph{IEEE Transactions on Signal Processing}, vol.~53, no.~8, pp.
  3010--3022, 2005.

\bibitem{mishra_spectral_2015}
K.~V. Mishra, M.~Cho, A.~Kruger, and W.~Xu, ``Spectral super-resolution with
  prior knowledge,'' \emph{IEEE Transactions on Signal Processing}, vol.~63,
  no.~20, pp. 5342--5357, 2015.

\bibitem{bhaskar_atomic_2013}
B.~N. Bhaskar, G.~Tang, and B.~Recht, ``Atomic norm denoising with applications
  to line spectral estimation,'' \emph{IEEE Transactions on Signal Processing},
  vol.~61, no.~23, pp. 5987--5999, 2013.

\bibitem{duval_exact_2015}
V.~Duval and G.~Peyr{\'e}, ``Exact support recovery for sparse spikes
  deconvolution,'' \emph{Foundations of Computational Mathematics}, vol.~15,
  no.~5, pp. 1315--1355, Oct. 2015.

\bibitem{chi_compressive_2015}
Y.~Chi and Y.~Chen, ``Compressive {Two}-{Dimensional} {Harmonic} {Retrieval}
  via {Atomic} {Norm} {Minimization},'' \emph{IEEE Transactions on Signal
  Processing}, vol.~63, no.~4, pp. 1030--1042, Feb. 2015.

\bibitem{li2015off}
Y.~Li and Y.~Chi, ``Off-the-grid line spectrum denoising and estimation with
  multiple measurement vectors,'' \emph{IEEE Transactions on Signal
  Processing}, vol.~64, no.~5, pp. 1257--1269, 2015.

\bibitem{yang_exact_2016}
Z.~Yang and L.~Xie, ``Exact joint sparse frequency recovery via optimization
  methods,'' \emph{IEEE Transactions on Signal Processing}, vol.~64, no.~19,
  pp. 5145--5157, 2016.

\bibitem{zhang2019efficient}
Z.~Zhang, Y.~Wang, and Z.~Tian, ``Efficient two-dimensional line spectrum
  estimation based on decoupled atomic norm minimization,'' \emph{Signal
  Processing}, vol. 163, pp. 95--106, 2019.

\bibitem{yang2015enhancing}
Z.~Yang and L.~Xie, ``Enhancing sparsity and resolution via reweighted atomic
  norm minimization,'' \emph{IEEE Transactions on Signal Processing}, vol.~64,
  no.~4, pp. 995--1006, 2015.

\bibitem{li2020approximate}
Q.~Li and G.~Tang, ``Approximate support recovery of atomic line spectral
  estimation: A tale of resolution and precision,'' \emph{Applied and
  Computational Harmonic Analysis}, vol.~48, no.~3, pp. 891--948, 2020.

\bibitem{duval2020characterization}
V.~Duval, ``A characterization of the non-degenerate source condition in
  super-resolution,'' \emph{Information and Inference: A Journal of the IMA},
  vol.~9, no.~1, pp. 235--269, 2020.

\bibitem{da2020stable}
M.~{Ferreira Da Costa} and Y.~{Chi}, ``On the stable resolution limit of total
  variation regularization for spike deconvolution,'' \emph{IEEE Transactions
  on Information Theory}, vol.~66, no.~11, pp. 7237--7252, 2020.

\bibitem{li2018atomic}
S.~Li, D.~Yang, G.~Tang, and M.~B. Wakin, ``Atomic norm minimization for modal
  analysis from random and compressed samples,'' \emph{IEEE Transactions on
  Signal Processing}, vol.~66, no.~7, pp. 1817--1831, 2018.

\bibitem{fu2018quantized}
H.~Fu and Y.~Chi, ``Quantized spectral compressed sensing: Cramer--rao bounds
  and recovery algorithms,'' \emph{IEEE Transactions on Signal Processing},
  vol.~66, no.~12, pp. 3268--3279, 2018.

\bibitem{heckel2016super}
R.~Heckel, V.~I. Morgenshtern, and M.~Soltanolkotabi, ``Super-resolution
  radar,'' \emph{Information and Inference: A Journal of the IMA}, vol.~5,
  no.~1, pp. 22--75, 2016.

\bibitem{heckel_generalized_2018}
R.~Heckel and M.~Soltanolkotabi, ``Generalized line spectral estimation via
  convex optimization,'' \emph{IEEE Transactions on Information Theory},
  vol.~64, no.~6, pp. 4001--4023, 2018.

\bibitem{wang2019super}
Y.~Wang, Y.~Zhang, Z.~Tian, G.~Leus, and G.~Zhang, ``Super-resolution channel
  estimation for arbitrary arrays in hybrid millimeter-wave massive {MIMO}
  systems,'' \emph{IEEE Journal of Selected Topics in Signal Processing},
  vol.~13, no.~5, pp. 947--960, 2019.

\bibitem{ongie2016off}
G.~Ongie and M.~Jacob, ``Off-the-grid recovery of piecewise constant images
  from few fourier samples,'' \emph{SIAM Journal on Imaging Sciences}, vol.~9,
  no.~3, pp. 1004--1041, 2016.

\bibitem{wei2020gridless}
Z.~{Wei}, W.~{Wang}, F.~{Dong}, and Q.~{Liu}, ``Gridless one-bit
  direction-of-arrival estimation via atomic norm denoising,'' \emph{IEEE
  Communications Letters}, vol.~24, no.~10, pp. 2177--2181, 2020.

\bibitem{chi_guaranteed_2016}
Y.~Chi, ``Guaranteed {Blind} {Sparse} {Spikes} {Deconvolution} via {Lifting}
  and {Convex} {Optimization},'' \emph{IEEE Journal of Selected Topics in
  Signal Processing}, vol.~10, no.~4, pp. 782--794, 2016.

\bibitem{li_stable_2019}
Y.~Li and Y.~Chi, ``Stable separation and super-resolution of mixture models,''
  \emph{Applied and Computational Harmonic Analysis}, vol.~46, no.~1, pp.
  1--39, 2019.

\bibitem{fernandez-granda_demixing_2017}
C.~Fernandez-Granda, G.~Tang, X.~Wang, and L.~Zheng, ``Demixing sines and
  spikes: {Robust} spectral super-resolution in the presence of outliers,''
  \emph{Information and Inference: A Journal of the IMA}, vol.~7, no.~1, pp.
  105--168, 2017.

\bibitem{ferreira2019self}
M.~{Ferreira Da Costa} and Y.~{Chi}, ``Self-calibrated super resolution,'' in
  \emph{2019 53rd Asilomar Conference on Signals, Systems, and Computers},
  2019, pp. 230--234.

\bibitem{zhu2012faster}
L.~Zhu, W.~Zhang, D.~Elnatan, and B.~Huang, ``Faster {STORM} using compressed
  sensing,'' \emph{Nature methods}, vol.~9, no.~7, pp. 721--723, 2012.

\bibitem{donoho_sparse_2005}
D.~L. Donoho and J.~Tanner, ``Sparse nonnegative solution of underdetermined
  linear equations by linear programming,'' \emph{Proceedings of the National
  Academy of Sciences}, vol. 102, no.~27, pp. 9446--9451, 2005.

\bibitem{denoyelle_support_2017}
Q.~Denoyelle, V.~Duval, and G.~Peyr{\'e}, ``Support recovery for sparse
  super-resolution of positive measures,'' \emph{Journal of Fourier Analysis
  and Applications}, vol.~23, no.~5, pp. 1153--1194, 2017.

\bibitem{morgenshtern_super-resolution_2016}
V.~I. Morgenshtern and E.~J. Candes, ``Super-resolution of positive sources:
  {The} discrete setup,'' \emph{SIAM Journal on Imaging Sciences}, vol.~9,
  no.~1, pp. 412--444, 2016.

\bibitem{schiebinger_superresolution_2017}
G.~Schiebinger, E.~Robeva, and B.~Recht, ``Superresolution without
  separation,'' \emph{Information and Inference: A Journal of the IMA}, vol.~7,
  no.~1, pp. 1--30, 2017.

\bibitem{morgenshtern2020super}
V.~I. Morgenshtern, ``Super-resolution of positive sources on an arbitrarily
  fine grid,'' \emph{arXiv preprint arXiv:2005.06756}, 2020.

\bibitem{eftekhari2019sparse}
A.~Eftekhari, J.~Tanner, A.~Thompson, B.~Toader, and H.~Tyagi, ``Sparse
  non-negative super-resolution -- simplified and stabilised,'' \emph{Applied
  and Computational Harmonic Analysis}, 2019.

\bibitem{pal2010nested}
P.~Pal and P.~P. Vaidyanathan, ``Nested arrays: A novel approach to array
  processing with enhanced degrees of freedom,'' \emph{IEEE Transactions on
  Signal Processing}, vol.~58, no.~8, pp. 4167--4181, 2010.

\bibitem{qin2015generalized}
S.~Qin, Y.~D. Zhang, and M.~G. Amin, ``Generalized coprime array configurations
  for direction-of-arrival estimation,'' \emph{IEEE Transactions on Signal
  Processing}, vol.~63, no.~6, pp. 1377--1390, 2015.

\bibitem{vaidyanathan2010sparse}
P.~P. Vaidyanathan and P.~Pal, ``Sparse sensing with co-prime samplers and
  arrays,'' \emph{IEEE Transactions on Signal Processing}, vol.~59, no.~2, pp.
  573--586, 2010.

\bibitem{qiao2019guaranteed}
H.~{Qiao} and P.~{Pal}, ``Guaranteed localization of more sources than sensors
  with finite snapshots in multiple measurement vector models using difference
  co-arrays,'' \emph{IEEE Transactions on Signal Processing}, vol.~67, no.~22,
  pp. 5715--5729, 2019.

\bibitem{tan2014direction}
Z.~Tan, Y.~C. Eldar, and A.~Nehorai, ``Direction of arrival estimation using
  co-prime arrays: A super resolution viewpoint,'' \emph{IEEE Transactions on
  Signal Processing}, vol.~62, no.~21, pp. 5565--5576, 2014.

\bibitem{ferreira_da_costa_low_2017}
M.~Ferreira Da~Costa and W.~Dai, ``Low dimensional atomic norm representations
  in line spectral estimation,'' in \emph{2017 {IEEE} {International}
  {Symposium} on {Information} {Theory} ({ISIT})}, Jun. 2017, pp. 226--230.

\bibitem{cosse2020compressed}
A.~Cosse, ``Compressed super-resolution {I}: Maximal rank sum-of-squares,''
  \emph{arXiv preprint arXiv:2001.01644}, 2020.

\bibitem{chandrasekaran_convex_2012}
V.~Chandrasekaran, B.~Recht, P.~A. Parrilo, and A.~S. Willsky, ``The convex
  geometry of linear inverse problems,'' \emph{Foundations of Computational
  Mathematics}, vol.~12, no.~6, pp. 805--849, 2012.

\bibitem{de_castro_exact_2012}
Y.~De~Castro and F.~Gamboa, ``Exact reconstruction using {Beurling} minimal
  extrapolation,'' \emph{Journal of Mathematical Analysis and Applications},
  vol. 395, no.~1, pp. 336--354, 2012.

\bibitem{ferreira_da_costa_tight_2018}
M.~Ferreira Da~Costa and W.~Dai, ``A tight converse to the spectral resolution
  limit via convex programming,'' in \emph{2018 {IEEE} {International}
  {Symposium} on {Information} {Theory} ({ISIT})}, Jun. 2018, pp. 901--905.

\bibitem{tang2015resolution}
G.~Tang, ``Resolution limits for atomic decompositions via {M}arkov-{B}ernstein
  type inequalities,'' in \emph{2015 International Conference on Sampling
  Theory and Applications (SampTA)}.\hskip 1em plus 0.5em minus 0.4em\relax
  IEEE, 2015, pp. 548--552.

\bibitem{fejer1916trigonometrische}
L.~Fej{\'e}r, ``{\"U}ber trigonometrische polynome.'' \emph{Journal f{\"u}r die
  reine und angewandte Mathematik}, vol. 1916, no. 146, pp. 53--82, 1916.

\bibitem{liu2017maximally}
C.-L. Liu and P.~Vaidyanathan, ``Maximally economic sparse arrays and {C}antor
  arrays,'' in \emph{2017 IEEE 7th International Workshop on Computational
  Advances in Multi-Sensor Adaptive Processing (CAMSAP)}.\hskip 1em plus 0.5em
  minus 0.4em\relax IEEE, 2017, pp. 1--5.

\bibitem{grant_cvx:_nodate}
M.~Grant and S.~Boyd, \emph{{CVX}: {Matlab} software for disciplined convex
  programming}.

\bibitem{chandrasekaran2013computational}
V.~Chandrasekaran and M.~I. Jordan, ``Computational and statistical tradeoffs
  via convex relaxation,'' \emph{Proceedings of the National Academy of
  Sciences}, vol. 110, no.~13, pp. E1181--E1190, 2013.

\end{thebibliography}
\bibliographystyle{IEEEtran}

\end{document}